\documentclass[universe,accept,pdftex,10pt,a4paper]{mdpi}
\firstpage{1} 
\articlenumber{43}
\doinum{10.3390/universe3020043}
\pubvolume{3}
\pubyear{2017}
\copyrightyear{2017}
\externaleditor{Academic Editor: Lorenzo Iorio}
\history{Published: 8 May 2017}

\Title{Mirror QCD and Cosmological Constant}

\Author{Roman Pasechnik$^{1}$, George Prokhorov$^2$ and Oleg Teryaev$^2$}
\AuthorNames{Roman Pasechnik, George Prokhorov, Oleg Teryaev}

\address{$^{1}$Department of Astronomy and Theoretical Physics, Lund University, SE-223 62 Lund, Sweden\\
\vspace{0.3cm}
$^{2}$Bogoliubov Laboratory of Theoretical Physics, 
Joint Institute for Nuclear Research, Dubna, Russian Federation}

\corres{Correspondence: Roman.Pasechnik@thep.lu.se}

\abstract{An analog of Quantum Chromo Dynamics (QCD) sector known as mirror QCD (mQCD) can affect the cosmological evolution 
due to a non-trivial contribution to the Cosmological Constant analogous to that induced by the ground state in non-perturbative QCD. 
In this work, we explore a plausible hypothesis for trace anomalies cancellation between the usual QCD and mQCD. Such an anomaly cancellation 
between the two gauge theories, if exists in Nature, would lead to a suppression or even elimination of their contributions to the Cosmological Constant. 
The trace anomaly compensation condition and the form of the non-perturbative mQCD coupling constant in the infrared limit have 
been proposed by analysing a partial non-perturbative solution of the Einstein--Yang-Mills equations of motion.}



\begin{document}

\section{Introduction}

The ground state of Yang-Mills (YM) theories plays a critical role in both Particle Physics and Cosmology. In particular,
the gluon condensate in Quantum Chromo Dynamics (QCD) largely determines non-trivial properties of the topological QCD vacuum and is responsible e.g. 
for the color confinement effects and hadron mass generation which can be understood beyond the Perturbation Theory (PT) only (for a comprehensive review 
on the QCD vacuum, see e.g. Refs.~\cite{Shifman:1978bx,Schafer:1996wv,Diakonov:2002fq,Diakonov:2009jq} and references therein). 
The gluon condensate directly influences properties of the quark-gluon plasma and its hadronisation, as well as dynamics of the QCD phase transition. 
On the other hand, YM condensates have various implications in the cosmological evolution ranging from the Cosmic Inflation 
\cite{Zhang:1994pm,Maleknejad:2011sq,Zhitnitsky:2013pna} to the phenomenon of late-time acceleration and the Dark Energy (DE) 
\cite{Thomas:2009uh,Pasechnik:2013poa,Dona:2015xia} (see also Refs.~\cite{Galtsov:1991un,Cavaglia:1993en,Bamba:2008xa,
Elizalde:2010xq,Galtsov:2011aa,Elizalde:2012yk}). 

Currently, the Cosmological Constant (CC) with the vacuum equation of state $w\equiv p/\epsilon=-1$ is a preferred scenario for the late-time acceleration epoch 
supported by a wealth of recent observations provided that $w=-1.006 \pm 0.045$ (see e.g. Refs.~\cite{WMAP,Planck}). Despite of many DE/CC models existing 
in the literature, there is not a compelling resolution of the CC problem i.e. why the CC term is small and positive as well as why the CC term is non-zeroth and exists at all
(for the existing proposals in connection to the hierarchy and origin problems of the Standard Model (SM) see e.g. Refs.~\cite{Bjorken:2001yv,Bass:2015csr,
Jegerlehner:2015cva}). From the Quantum Field Theory (QFT) viewpoint, the ground state energy density of the universe should account for a bulk of various 
contributions from existing quantum fields at energy scales ranging from the Quantum Gravity (Planck) scale, $M_{\rm PL}\simeq 1.2\cdot 10^{19}$ GeV, down 
to the QCD confinement scale, $\Lambda_{\rm QCD}\simeq 0.1$ GeV. Of course, in absence of a workable version of quantum gravity the CC problem is not fully 
defined yet so in what follows non-quantum-gravitational aspects of this problem. Indeed, even such relatively well-known vacuum subsystems of the SM as the Higgs 
and quark-gluon condensates (which supposedly have nothing to do with quantum gravity) exceed by far the observed cosmological constant which is often 
considered as a severe problem \cite{Weinberg:1988cp,Wilczek:1983as} (for recent reviews on this topic, see e.g. Refs.~\cite{Pasechnik,our-review,Sola:2013gha} 
and references therein). Also it is well known, that an every field in the universe forms a divergent perturbative vacuum contribution, which is usually cut off at 
the Planck scale. The cancellation of these contributions may need the introduction of additional bosonic and fermionic fields putting important constraints for 
the particle spectrum \cite{Kamenshchik:2006nm,Alberghi:2008zz,Kamenshchik:2016tjz}.

In this work, we discuss a well-defined part of the CC problem connected with formation of big non-perturbative vacuum contributions on the hadronic scale after 
the QCD phase transition, assuming that the contributions from higher scales are already compensated. In the case of confined QCD with color $SU(3)$ gauge 
symmetry, there is a rather unique (negative-valued) contribution to the ground state energy of the universe originating from the non-perturbative quantum 
fluctuations of the quark and gluon fields \cite{Boucaud:2002nc,Hutter:1995sc,Shifman:1978bx,Schafer:1996wv}, $\epsilon^{\rm QCD}<0$. Given the fact 
that the CC term observed in astrophysical measurements is very small (and positive-valued),
\begin{equation}
\label{Lcosm-ratio}
\epsilon_{\rm CC}>0\,, \qquad \Big| \frac{\epsilon_{\rm CC}}{\epsilon^{\rm QCD}}\Big| \simeq 10^{-44} \,,
\end{equation}
one must eliminate the QCD vacuum contribution, $\epsilon^{\rm QCD}$, with an unprecedented accuracy over forty decimal digits.
A dynamical mechanism for such a gross cancellation of vacua terms is yet theoretically unknown although several possible 
scenarios elaborating on the cosmological role of the QCD condensates have been discussed so far e.g. in Refs.~\cite{Pasechnik,Copeland:2006wr,Dolgov,
Brodsky:2009zd,Pasechnik:2013poa,Pasechnik:2013sga}. This work is devoted to making a further step in exploring the possibility for vacua cancellations 
in quantum YM theories with a non-trivial ground state. Clearly, in order to cancel the QCD vacuum contribution, $\epsilon^{\rm QCD}$, formed during 
the QCD phase transition epoch, a positive contribution to the vacuum energy density should be formed at the same QCD energy scale $\Lambda_{\rm QCD}$. 
Where could such an extra contribution originate from? 

Here we suggest a new scenario of compensation realized by means of a hidden (mirror) sector of particles \cite{Strassler:2006im} which correspond 
to the extra non-Abelian gauge group and whose possible interaction with the visible SM sectors is strongly suppressed.  

In particular, a class of models known as Neutral Naturalness theories has been proposed in the literature \cite{Chacko:2005pe,Burdman:2006tz,Cai:2008au} 
as a promising solution of the naturalness problem in the SM protecting the weak scale from large radiative corrections. Various phenomenological implications 
of such a ``Mirror World'' concept have been discussed e.g. in Ref.~\cite{Barbieri:2005ri}. In particular, the mirror color $SU(3)$ gauge group is typically 
assumed to be a symmetry describing the confined phase in full analogy with ordinary QCD revealing interesting signatures at the Large Hadron Collider 
due to e.g. a mixing of mirror glueballs with the Higgs boson \cite{Chacko:2015fbc}.

Quite naturally, the quantum vacua contributions from the ``Mirror World'' should contribute to the CC on the same footing as known vacua since ``mirror''
particles are expected to gravitate in the same way as the usual ones. We argue that mirror QCD (mQCD) sector, if exists, should affect the cosmological expansion, 
in particular, via an extra non-trivial ``mirror gluon'' contribution into the ground state energy of the universe. In particular, an invisible QCD sector reconstructing 
Dark Energy was suggested in Refs.~\cite{Dona:2015xia, Alexander:2016xbm} while a unified description of Dark Matter and Dark Energy originating from invisible 
QCD dynamics was proposed in Refs.~\cite{Addazi:2016sot, Addazi:2016nok}. Below, we will demonstrate that under certain conditions the mirror gluon condensate 
can contribute to the energy density of the universe with positive sign and thus may, in principle, eliminate the negative QCD vacuum 
effect yielding a vanishingly small CC term. Attributing the positive vacuum energy contribution to the mQCD sector non-interacting with quarks and 
gluons in ordinary QCD, one may therefore resolve the issue of why such a positive-valued condensate energy density does not emerge in QCD hadron physics and 
affects the CC-term only. Then the observed CC can, in principle, be formed as a remnant of the gluon condensate cancellation in expanding universe (e.g. due to 
an uncompensated quantum gravity correction to the QCD ground state energy) \cite{Pasechnik:2013poa}, which appears to be remarkably consistent with 
the observed CC and with the Zeldovich scaling relation \cite{Zeldovich:1967gd}. The exact compensation of the QCD vacuum effect by means of the mirror 
gluon condensate is therefore the central point to the observable smallness of the CC.

\section{QCD and mirror QCD vacua compensation}
\label{Sec:compensation}

The condensate in QCD is formed by the contributions of gluon and quark non-perturbative quantum fluctuations
\begin{eqnarray}
\label{Linst-0}
\epsilon^{\rm QCD} =\epsilon^{\rm g}+\epsilon^{\rm q}\simeq -(5\pm 1)\times 10^{9}\;\text{MeV}^4 \,,
\qquad
\epsilon^{\rm q}=\frac{1}{4}\langle 0|m_u \bar{u}u + m_d\bar{d}d + m_s\bar{s}s|0\rangle\,.
\end{eqnarray} 
Usually, the dominant gluon contribution is given by means of the trace anomaly relation in QCD
\begin{eqnarray}
\label{Linst-1}
\epsilon^{\rm g} \equiv \frac{1}{4}\,\langle 0| T^{\mu,{\rm g}}_{\mu}  |0 \rangle \,, \qquad 
T_{\mu}^{\mu,{\rm g}}=\frac{\beta({\bar g}_s^2)}{2}\,F_{\mu\nu}^aF^{\mu\nu}_a \,,
\end{eqnarray} 
which to one-loop order reads \cite{Shifman:1978bx,Schafer:1996wv}
\begin{eqnarray}
\label{Linst-2}
\epsilon^{\rm g} = -\frac{b}{32}\langle0|\frac{\alpha_s}{\pi}F^a_{\mu\nu}
F_a^{\mu\nu}|0\rangle \,, \qquad \alpha_s=\frac{{\bar g}_s^2}{4\pi} \,,
\end{eqnarray} 
where $b=9$ is the first (one-loop) coefficient of the negative perturbative $\beta$-function 
in $SU(3)$ gluodynamics with three light flavors 
\begin{eqnarray}
\beta({\bar g}_s^2)=-\frac{b{\bar g}_s^2}{16\pi^2}<0\,, \qquad {\bar g}_s^2=\frac{16\pi^2}{\displaystyle b\ln(Q^2/\Lambda_{\rm QCD}^2)}\,,
\label{beta}
\end{eqnarray} 
with $\Lambda_{\rm QCD}$ being the QCD scale parameter (the normalization of $\beta$ corresponds to Ref.~\cite{Voloshin}). 
The formation of the chromomagnetic gluon condensate $\langle F^2 \rangle>0$ is typically considered at characteristic momentum 
scales $\mu_g$ inverse to the correlation length $l_g$, i.e. $\mu_g\sim l_g^{-1}\simeq 1.2$ GeV \cite{Shifman:1978bx,Schafer:1996wv}, 
where the perturbative QCD still provides a realistic estimate. This validates the use of one-loop approximated expression (\ref{Linst-2}).

One would like to explain why such a big negative contribution (\ref{Linst-0}), which is responsible for a variety of well-known 
phenomena in hadron physics and is rather unique for QCD, does not affect the cosmological expansion at late times. 
Provided that the observed CC-term density
\begin{equation}
\label{Lcosm}
\epsilon_{\rm CC}\simeq 3\times 10^{-35}\,{\rm MeV}^4 \,,
\end{equation}
is tiny compared to the QCD vacuum density (\ref{Linst-2}), the latter should be almost totally eliminated during 
the QCD phase transition epoch. Which mechanism could be responsible for that?

A mirror copy of QCD may generate a similar gluon contribution to the trace anomaly proportional to the corresponding 
$\beta$-function in mQCD,
\begin{eqnarray}
\epsilon^{\rm mQCD}_{\rm gluon}\equiv \frac{1}{4}\,\langle 0| T^{\mu,{\rm mQCD}}_{\mu}  |0 \rangle \propto \beta({\bar g}^2) \,.
\end{eqnarray}
In mQCD framework mirror quarks can be much heavier than in ordinary QCD \cite{Cai:2008au}. Applying the idea, that mQCD is similar in main features 
to usual QCD, this means, that in mQCD the vacuum is formed only by mirror gluon contribution with pure gluonic $\beta$-function, as long as 
the heavy quark condensates \cite{Generalis:1983hb} are compensated by quark part of $\beta$-function, i.e.
\begin{eqnarray}
\epsilon^{\rm mQCD}=\epsilon^{\rm mQCD}_{\rm gluon}\,.
\end{eqnarray}
  A possible cancellation of QCD and mQCD vacuum may ensure a required smallness of the observable CC density
\begin{eqnarray}
\epsilon^{\rm QCD} \simeq -\epsilon^{\rm mQCD} \,, 
\label{compC-1}
\end{eqnarray}
which means that the corresponding mirror gluon condensate should provide a positive contribution to the vacuum 
density, i.e. $\epsilon^{\rm mQCD}>0$. We suppose, that mQCD gluon condensate can compensate both gluon 
and quark condensates of usual QCD.

Adopting the traditional hypothesis that the mQCD sector of mirror quarks and gluons is confined but is not (or very weakly) 
interacting with the observed SM sectors \cite{Chacko:2015fbc} and considering only chromomagnetic condensates, 
the compensation condition (\ref{compC-1}) can be satisfied if and only if the mQCD $\beta$-function is positive, i.e. 
\begin{eqnarray}
\epsilon^{\rm mQCD}>0 \,, \qquad \langle F_{\rm mQCD}^2 \rangle>0 \,, \qquad \beta({\bar g}^2)>0 \,,
\label{compC-2}
\end{eqnarray}
which is not realized in the perturbative mQCD regime due to Eq.~(\ref{beta}). It is, however, possible to achieve 
the positivity of the non-perturbative $\beta$-function provided that at the characteristic scale
of the QCD gluon condensate formation, $\mu_g$, the mQCD sector is in deeply non-perturbative
regime. The latter condition can be satisfied if the mQCD scale parameter is large, i.e.
\begin{eqnarray}
\Lambda_{\rm mQCD} \gg \mu_g \simeq 1.2\; \rm{GeV} \,,
\label{Lam-mQCD}
\end{eqnarray}
such that the mirror gluon condensate would be in deeply non-perturbative regime by the moment in 
the cosmological evolution when its density gets precisely cancelled with the QCD contribution.
Note that the compensation conditions for the QCD and mQCD contributions (\ref{compC-1}) and 
(\ref{compC-2}), if indeed realized in nature, may be one of the most important implications of the mirror QCD 
in Cosmology yielding a vanishing CC-term and thus providing a dynamical way to resolve the CC problem.

Can the sign of the $\beta$-function in mQCD become positive in the non-perturbative regime?
In order to answer this question, one has to employ a proper formalism which extends the effective
action approach in a gauge theory beyond the perturbativity domain. Indeed, the compensation 
(\ref{compC-1}) emerges as a long-distance phenomenon and thus should hold beyond the PT.

\section{Effective Yang-Mills theory in expanding universe}
\label{Sec:eff-YM}

The effective action of the quantum YM gauge $SU(N)$($N=2,3,\dots$) theory consistently incorporating the vacuum polarisation effects 
and leading to the trace anomaly can be properly generalised to the FLRW background as follows \cite{Savvidy,Pagels} 
(see also Ref.~\cite{Dona:2015xia})
\begin{eqnarray} \nonumber
&& S_{\rm eff}[\mathcal{A}] = \int \mathcal{L}_{\rm eff}\sqrt{-g} d^4x \,, \qquad \mathcal{L}_{\rm
eff}=\frac{J}{4{\bar g}^2(J)}\,, \qquad J=-\frac{\mathcal{F}^2}{\sqrt{-g}}\,, 
\qquad \mathcal{F}^2\equiv \mathcal{F}^a_{\mu\nu}\mathcal{F}_a^{\mu\nu}\,, \\
&& g\equiv {\rm det}(g_{\mu\nu}) \,, \qquad g_{\mu\nu}=a(\eta)^2\mathrm{diag}(1,\,-1,\,-1,\,-1)\,, \qquad t = \int a(\eta) d\eta \,,
\label{Lrg}
\end{eqnarray}
where the YM field and the corresponding stress tensor are defined as usual
\begin{equation*}
\mathcal{A}_\mu^a\equiv {\bar g}\,A_\mu^a\,, \qquad \mathcal{F}^a_{\mu\nu}\equiv {\bar g}\,F^a_{\mu\nu}\,, \qquad
F^a_{\mu\nu}=\partial_\mu A^a_\nu - \partial_\nu A^a_\mu + {\bar g}\,f^{abc} A^b_\mu A^c_\nu
\end{equation*}
with internal (in adjoint representation) $a,b,c=1,\dots\, N^2-1$ and Lorentz $\mu,\nu=0,1,2,3$ indices and 
the gauge coupling ${\bar g}={\bar g}(J)$ satisfying the RG evolution equation \cite{Savvidy,Pagels}
\begin{equation}
\displaystyle 2J\frac{d{\bar g}^2}{dJ}={\bar g}^2\,\beta({\bar g}^2) \,.
\label{rg}
\end{equation}
Depending on the sign of the invariant $J$, one distinguishes the chromoelectric $J>0$ and chromomagnetic $J<0$ YM fields.  

The effective YM equation of motion in a non-trivial background metric reads
\begin{eqnarray} 
\label{YMeq}
\left(\frac{\delta^{ab}}{\sqrt{-g}}\partial_\nu\sqrt{-g}-f^{abc}\mathcal{A}_\nu^c\right)
\left[\frac{\mathcal{F}_b^{\mu\nu}}{{\bar g}^2\,\sqrt{-g}}\,
\Big(1-\frac12\beta\big({\bar g}^2\big)\Big)\right]=0 \,.
\end{eqnarray}
and can be employed beyond the PT as long as the non-perturbative $\beta$-function is known.
Note, this equation is the {\it exact} equation of the Einstein-YM theory as it does not imply any 
approximations. Remarkably enough, this equation has a simple manifestly non-perturbative and exact 
ground-state solution with positive $\beta$-function
\begin{eqnarray}
\label{SS}
\beta\big({\bar g}^2(J)\big) = 2 \,,
\end{eqnarray}
which is a complete analog of similar solution (see Ref.~\cite{Pagels}, Eqs.~(13) and (17)) found in the Euclidian case with 
the negative coupling and the $\beta$-function corresponding to the ferromagnetic vacuum. It is also a non-perturbative 
analog of the perturbative solution~\cite{Zhang:1994pm,Pasechnik:2013sga} eliminating the traceless part 
of the energy-momentum tensor
\begin{eqnarray}\nonumber
T^{\nu}_{\mu}&=&\frac{1}{{\bar g}^2}\Big[1-\frac12\beta({\bar g}^2)\Big]
\Big(-\frac{\mathcal{F}^a_{\mu\lambda}\mathcal{F}_a^{\nu\lambda}}{\sqrt{-g}}-\frac{1}{4}\delta_{\mu}^{\nu}\,J\Big) -
\frac{\delta_{\mu}^{\nu}\beta({\bar g}^2)}{8{\bar g}^2}\,J \,,
\label{T}
\end{eqnarray}
which then takes the following form
\begin{eqnarray}
T^{\nu}_{\mu,0}&=&-\frac{J}{4{\bar g}^2(J)}\delta^{\nu}_{\mu}\,.
\label{T1}
\end{eqnarray}
We have explicitly checked that the exact partial solution (\ref{SS}) of the YM equation (\ref{YMeq}) naturally corresponds to the minimum of 
the non-perturbative effective YM Lagrangian (\ref{Lrg}). It holds strictly beyond the Perturbation Theory, just like the YM trace anomaly itself. 
Thus, the exact solution (\ref{SS}) corresponds to the physical quantum ground state of an effective YM theory. 

It is important to point out following Ref.~\cite{Pagels}, that the equations (\ref{SS}-\ref{T1}) were obtained in the pure YM case when the interaction 
with other fields can be neglected. In particular, we neglect the mQCD quark current in the right hand side of Eq.~(\ref{YMeq}). Let us also stress 
that we consider the effective YM Lagrangian (\ref{Lrg}) and energy-momentum tensor (\ref{T}) as a classical model \cite{Pagels} 
which possesses well-known properties of the full quantum theory such as (i) local gauge invariance, (ii) RG evolution and asymptotic 
freedom, (iii) correct quantum vacuum configurations, and (iv) trace anomaly given by the last term in Eq.~(\ref{T}). These provide 
a sufficient motivation and physics interest in cosmological aspects of the considering effective model. 

As the solution (\ref{SS}) leads to the energy-momentum tensor of vacuum type (\ref{T1}), it immediately follows from the Friedmann 
equations that the corresponding energy density is constant
\begin{eqnarray}
-\frac{J}{4{\bar g}^2(J)}=\rm const\,,
\label{const}
\end{eqnarray}
thus, the contribution of the YM fields has a cosmological constant form. In particular, this can be realised if $J=\rm const$. 
Indeed, the solution (\ref{SS}) fixes the invariant $J$ to its constant initial value 
\begin{eqnarray}
J(t)\equiv J(t=0)=J_0\,.
\label{J}
\end{eqnarray}
Such solutions were also considered in Refs.~\cite{Pagels,Zhang:1994pm,Pasechnik:2013sga} in connection with the spontaneous vacuum 
magnetisation and in the domain concept of the QCD vacuum \cite{Nachtmann:1983uz} (see also the recent paper \cite{Nedelko:2016gdk} and 
references therein).

\begin{figure}[!h]
 \centerline{\includegraphics[width=0.45\textwidth]{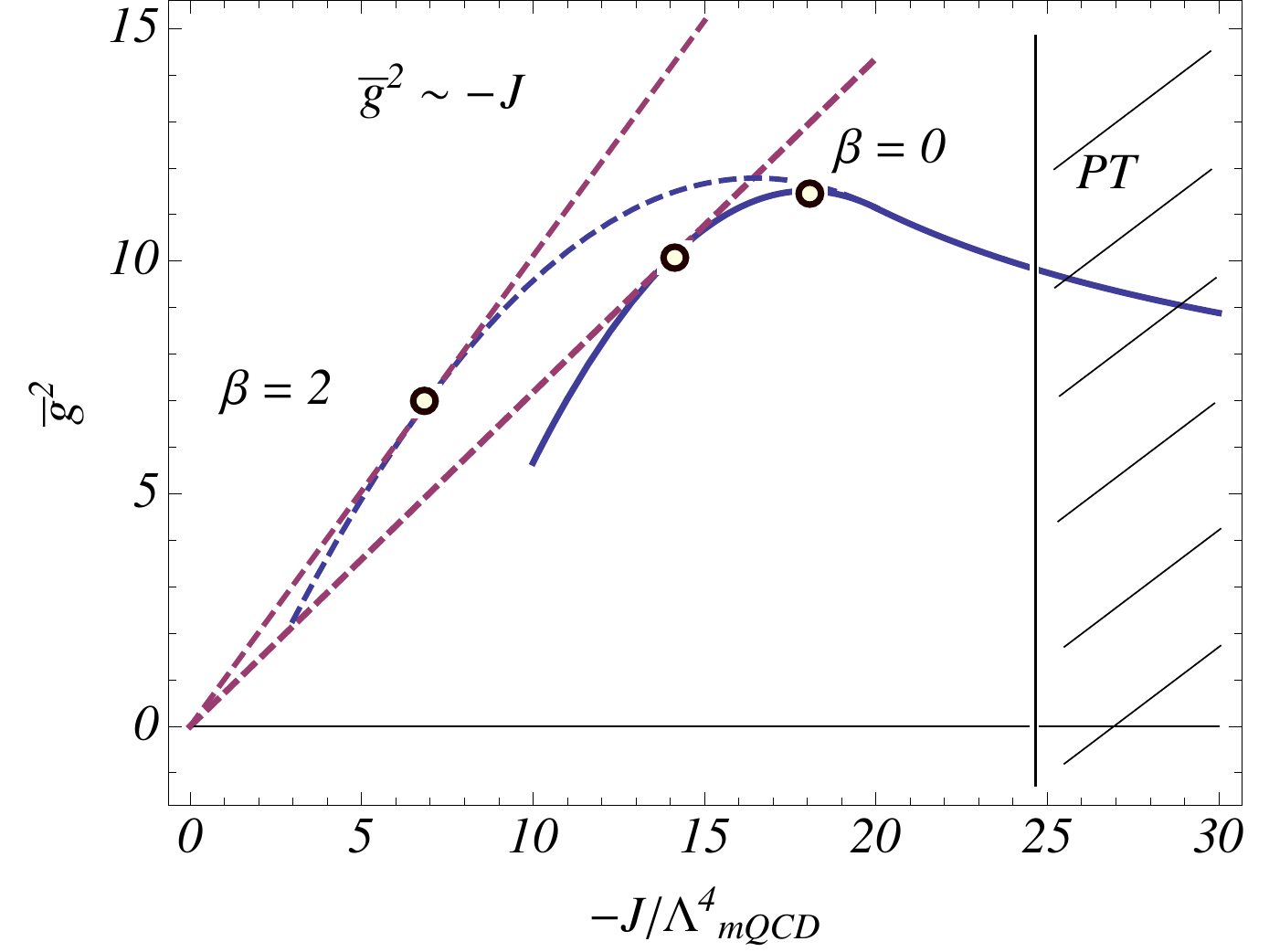}}
   \caption{An example of the non-perturbative mQCD coupling constant ${\bar g}^2={\bar g}^2(J)$ 
   behavior as a function of $J$ in consistency with 
   the non-perturbative solution found in Eq.~(\ref{SS}).}
   \label{fig:g2}
\end{figure}

Further, we will apply the solution (\ref{SS}) and (\ref{J}) to the the mQCD theory in the non-perturbative regime. The energy-momentum 
tensor in this case becomes constant as expected
\begin{eqnarray}
T^{\nu*}_{\mu}=\epsilon^{\rm mQCD}\delta_{\mu}^{\nu}\,, \qquad \epsilon^{\rm mQCD}\equiv 
-\frac{J^{\rm mQCD}_0}{4{\bar g}^2_0}\,,
\label{Tsol1}
\end{eqnarray}
where $\bar{g}^2_0=\bar{g}^2(J_0)$. The coupling $\bar{g}^2(J)$ touches the linear function $f(J)=\bar{g}^2_0\cdot (J/J_0)$ 
at the point $J_0$ (indeed, it has the same value and derivative). And vice versa, if $\bar{g}^2(J)$ touches ${\rm const}\cdot J$ 
at some point $J_0$, then $\frac{d\bar{g}^2}{dJ}|_{J=J_0}=\frac{\bar{g}^2_0}{J_0}$, which means that 
Eq.~(\ref{SS}) is satisfied at $J_0$. Indeed, the existence of such a contact point is a necessary and sufficient condition for 
the solution (\ref{SS}) with fixed $J=J_0$. This allows us to constrain generic non-perturbative behavior of the corresponding 
$\bar{g}^2(J)$. An illustration of the corresponding infrared behavior of the mQCD coupling in consistency with both the non-perturbative 
asymptotics for the $\beta$-function (\ref{SS}) and the conventional perturbative regime of asymptotic freedom (\ref{beta}) 
is shown in Fig.~\ref{fig:g2}. A desirable non-monotonic shape of the coupling was earlier discussed in the case of usual QCD 
\cite{Baldicchi:2007ic, Shirkov:2008hf}.

From Eq.~(\ref{const}) one notices that the mQCD gauge field gives a constant vacuum contribution to the energy-momentum 
tensor in the Einstein equations. Since the QCD and mQCD contributions to the ground-state energy density have opposite signs 
there is a compelling possibility that they can, in principle, cancel each other at some moment $t=t^*$ in the cosmological history provided that
\begin{eqnarray}
\label{compens}
\epsilon^{\rm mQCD}\to -\epsilon^{\rm QCD}
\end{eqnarray}
in the infrared regime of mQCD. We argue that this relation can be reached at some $t=t^*$ around the QCD phase transition epoch
as long as the mQCD energy scale $\Lambda_{\rm mQCD}$ is much larger than that in QCD $\Lambda_{\rm QCD}$ such that at 
the cancellation time $t^*$ their $\beta$-functions have different signs. Keeping ${\bar g}_0^2>0$ in both QCD and mQCD, 
we arrive at the following form of the compensation condition (\ref{compC-1})
\begin{eqnarray}
\label{compens-1}
\epsilon^{\rm QCD} \simeq \frac{J^{\rm mQCD}_0}{4{\bar g}^2_0} <0\,,\qquad J^{\rm mQCD}_0<0\, ,
\end{eqnarray}
which means, that mQCD condensate has to be chromomagnetic.

After such a compensation is achieved, only a very small $\epsilon_{\rm CC}$ contribution, which could be formed 
by other vacuum sources $\epsilon_{\rm vac}$ and possibly by a non-compensated part of mQCD and QCD vacua 
contributions $\epsilon^{\rm QCD} + \epsilon^{\rm mQCD}$, remains. Of course, a fine-tuning is unavoidable 
to match the observations, although would not be entirely unreasonable due to the same order magnitude of 
the QCD and mQCD contributions. Such a vacua alignment, if realised in Nature, can be suggested e.g. by macroscopic 
cancellation of trace anomalies in gauge theories inspired by the phenomenologically well-known color confinement phenomenon. 
Then the standard Friedmann equation in the non-stationary 
FLRW universe
\begin{eqnarray}
 \frac{3}{\kappa}\frac{(a')^2}{a^4} =\epsilon_{\rm mat} + \epsilon_{\rm CC}\,, \qquad 
 \epsilon_{\rm CC} \equiv \epsilon^{\rm QCD} + \epsilon^{\rm mQCD} + \epsilon_{\rm vac} \,,
\end{eqnarray}
determines the cosmological evolution, $a=a(\eta)$, driven by the gluon and mirror gluon condensate 
densities (compensating each other exactly or in part), the matter contribution, $\epsilon_{\rm mat}$, and 
other possible vacua contributions of a different kind, $\epsilon_{\rm vac}$. 

Let us notice for completeness, that Eq.~(\ref{SS}) allows for a more specific solution apart from constant fields (\ref{J}). 
This another solution appears, if the non-perturbative $\beta$-function becomes constant satisfying Eq.~(\ref{SS}) in some finite 
domain which corresponds to the strong coupling regime (saturated behaviour). In this case, Eq.~(\ref{SS}) can be substituted 
into the RG equation
\begin{eqnarray}
\frac{d\ln {\bar g}^2}{d\ln\big(-J/(\xi\Lambda_{\rm mQCD})^4\big)}=\frac12 \beta\big({\bar g}^2\big) = 1 \,,
\label{rgNP}
\end{eqnarray}
($\xi$ is a numerical parameter) which implies that the mQCD coupling is proportional to $J$ in the infrared limit, e.g.
\begin{eqnarray}
{\bar g}^2(J)={\bar g}^2_0\,\frac{J}{J_0}\,, \qquad  {\bar g}^2_0\equiv {\bar g}^2(J_0) \,.
\label{J0}
\end{eqnarray}
Such behaviour clearly guarantees a constant vacuum energy (see Eq.~(\ref{T1})) as well as a possibility for the QCD/mQCD vacua 
compensation. According to Eq.~(\ref{J0}) the gauge coupling squared has to approach the linear ${\bar g}^2\sim J$ asymptotics 
in the non-perturbative regime (note, for the constant field solution (\ref{J}) ${\bar g}^2$ has to satisfy a much less restrictive 
constraint and just touches the linear asymptotics ${\bar g}^2\sim J$ at a fixed point). 

So, within the QCD/mQCD vacua compensation hypothesis, both vacua subsystems should be generated in early universe at close (but different) 
energy scales and then get compensated during the cosmological QCD phase transition epoch. As was shown above, this can be realised in a deeply 
non-perturbative regime for the mirror gluon condensate which asymptotically acquires the same absolute value of energy density and opposite sign 
compared to the QCD gluon one, such that they almost exactly eliminate each other at macroscopically large space-time separations.

\section{Summary}
\label{Sec:Summary}

The only non-perturbative strongly-coupled vacuum system known in the Standard Model is the QCD ground state. One of the basic aspects of 
the Cosmological Constant problem related to the fact that neither perturbative (e.g. Higgs) nor non-perturbative (such as QCD) quantum vacua affect 
the cosmological expansion remains unexplored. While perturbative (weakly-coupled) quantum vacua should be treated elsewhere e.g. in a proper theory 
of Quantum Gravity etc, the low-energy strongly-coupled vacua such that in QCD are the most problematic ones since they do not affect the Cosmological Constant 
but {\it a priori} they have nothing to do with quantum dynamics at the Planck scale. The absence of vacuum contributions of the color (quark and gluon) fields 
at macroscopic distances in QCD is tightly related with the confinement phenomenon, i.e. with the fact that no colored particles can propagate through 
macroscopic spacetime separations.

Note, we do not attempt to resolve the Cosmological Constant problem but consider a small but essential part of it connected with the QCD vacuum 
(provided that all perturbative vacua are eliminated by some other mechanism, see e.g. Ref.~\cite{Kamenshchik:2006nm,Alberghi:2008zz,Kamenshchik:2016tjz}).
Namely, in this Letter, we discuss a plausible hypothesis of a partial or exact cancellation of averaged vacuum densities between QCD and a mirror high-scale copy 
of QCD in the confined regime corresponding to large separations (while locally such a compensation may not hold).

In this work, we do not propose any dynamical mechanism for such a compensation between QCD and mirror QCD vacua as this task would require a more detailed 
knowledge of their real-time dynamics, unavailable at the current state of this research field. Instead, for the first time we derive simple and generic conditions 
under which their contributions to the Cosmological Constant are mutually eliminated, if at all realised in Nature (for earlier works on this topic, see 
Refs.~\cite{Pasechnik:2013sga,Pasechnik:2013poa}). These conditions are based upon our main new finding that the $\beta$-function of an effective 
Yang-Mills theory has opposite signs in its perturbative and non-perturbative regimes. By an appropriate fine-tuning of QCD and mirror QCD vacua parameters whose
dynamical reasons are unknown, the compensation can be provided by a partial non-perturbative solution of the Yang-Mills equation of motion corresponding to 
a positive constant $\beta$-function in deeply infrared regime of mirror QCD. 

{\it Acknowledgments}

We are indebted to A. Kamenshchik for useful comments. 
This work was partially supported by RFBR Grant 14-01-00647 
and by the Swedish Research Council, contract number 621-2013-428.


\end{document}